\documentclass[superscriptaddress,preprint]{revtex4-1}
\usepackage{amsmath,amssymb,bm}
\usepackage{graphicx}
\usepackage{dcolumn}
\usepackage{bm}
\usepackage{multirow}
\usepackage{color}
\begin{document}
\newcommand{\hc}{H$_{c2}$}
\newcommand{\cso}{Cu$_{2}$OSeO$_{3}$}

\title{Linearly polarized GHz magnetization dynamics of spin helix modes in the ferrimagnetic insulator \cso}

\author{I.~Stasinopoulos}
\affiliation{Physik Department E10, Technische Universit\"at M\"unchen, 85748 Garching, Germany}

\author{S.~Weichselbaumer}
\affiliation{Physik Department E10, Technische Universit\"at M\"unchen, 85748 Garching, Germany}

\author{A.~Bauer}
\affiliation{Physik Department E51, Technische Universit\"at M\"unchen, 85748 Garching, Germany}

\author{J.~Waizner}
\affiliation{Institut f\"ur Theoretische Physik, Universit\"at zu K\"oln, 50937 K\"oln, Germany}

\author{H.~Berger}
\affiliation{Institut de Physique de la Mati\`ere Complexe, \'Ecole Polytechnique F\'ed\'erale de Lausanne, 1015 Lausanne, Switzerland}

\author{M.~Garst}
\affiliation{Institut f\"ur Theoretische Physik, Universit\"at zu K\"oln, 50937 K\"oln, Germany}

\affiliation{Institut f\"ur Theoretische Physik, Technische Universit\"at Dresden, 01062 Dresden, Germany}

\author{C.~Pfleiderer}
\affiliation{Physik Department E51, Technische Universit\"at M\"unchen, 85748 Garching, Germany}

\author{D.~Grundler}\thanks{Electronic mail: dirk.grundler@epfl.ch}
\affiliation{Institute of Materials and Laboratory of Nanoscale Magnetic Materials and Magnonics (LMGN), \'Ecole Polytechnique F\'ed\'erale de Lausanne (EPFL), Station 12, 1015 Lausanne, Switzerland}

\date{\today}

\begin{abstract}
Linear dichroism \textemdash the polarization dependent absorption of electromagnetic waves\textemdash\ is routinely exploited in applications as diverse as structure determination of DNA or polarization filters in optical technologies. Here filamentary absorbers with a large length-to-width ratio are a prerequisite. For magnetization dynamics in the few GHz frequency regime strictly linear dichroism was not observed for more than eight decades. Here, we show that the bulk chiral magnet \cso\ exhibits linearly polarized magnetization dynamics at an unexpectedly small frequency of about 2\,GHz. Unlike optical filters that are assembled from filamentary absorbers, the magnet provides linear polarization as a bulk material for an extremely wide range of length-to-width ratios. In addition, the polarization plane of a given mode can be switched by 90$^\circ$ via a tiny variation in width. Our findings shed a new light on magnetization dynamics in that ferrimagnetic ordering combined with anisotropic exchange interaction offers strictly linear polarization and cross-polarized modes for a broad spectrum of sample shapes. The discovery allows for novel design rules and optimization of microwave-to-magnon transduction in emerging microwave technologies.
\end{abstract}


\maketitle

Microwave components such as power limiters, oscillators and tunable bandpass filters exploit the precession of magnetic moments (spins). The magnetization dynamics allows one to process microwave signals in the few GHz frequency regime. In microwave filters bulk spheres prepared from the ferrimagnetic insulator yttrium iron garnet (YIG) provide for instance an excellent quality factor and a wide tunability of frequencies. In microwave circulators and isolators the nonreciprocal characteristics of magnetic materials are functionalized. For an efficient operation, the magnetization dynamics and the microwaves require a common polarization. However, this is usually not the case as the equation of motion for spin precession (Landau-Lifshitz equation) provides a circular polarization\cite{LanLif1935,Gurevich}. It may be elliptically deformed by demagnetization fields\cite{LanLif1935,Gurevich,Gruenberg::NobelLecture}: For very thin films \cite{Zivieri2002} and filamentary magnetic rods with extremely large length-to-diameter ratios the ellipticity $\epsilon$ can take values close to one, but strictly linear polarization ($\epsilon=1$) is not possible (Supplementary Eqs. (1)-(3)). The sphere-like and bulk samples used so far in microwave components are neither filamentary nor in thin-film form. Hence their magnetization dynamics exhibit a polarization close to circular ($\epsilon=0$). In contrast, microwaves provided by standard coaxial cables, coplanar waveguides (CPWs) and microwave cavities are linearly polarized ($\epsilon=1$). As a consequence, 50 \% of the microwave energy would be wasted in case of a YIG sphere and nonreciprocal devices do not function due to the mismatch of polarization. This is a drawback for future on-chip resonators exploiting coupled magnons and photons for quantum information processing \cite{Huebl2013,Tabuchi2014,Zhang2014,Goryachev2014}. On the one hand, involved waveguides and cavity designs were  invented to produce microwaves with a high degree of circular polarization \cite{Wen1969,Duncan1957,Poole67,Yasukawa2016}. On the other hand magnetocrystalline anisotropy (e.g. easy-plane anisotropy) and antiferromagnetism \cite{Gurevich} might help to overcome the bad coupling with linearly polarized electromagnetic waves as these two qualities enhance the ellipticity in bulk materials. But at the same time, the frequency increases to several 10 GHz or even near-infrared frequencies (terahertz) \cite{Sievers::PhysRev::1963,Klausen2004,Kampfrath08} where efficient waveguides do not exist. The integration of magnets with microwave electronics could be significantly improved by a ferrimagnetic insulator similar to YIG that possesses linearly polarized magnetization dynamics. However, such a material and the underlying blueprint have not yet been identified.

Here, we report the discovery of linearly polarized magnetization dynamics in the chiral ferrimagnet \cso. This magnet supports a helical spin order stabilized by anisotropic exchange interaction, i.e., Dzyaloshinskii-Moriya interaction (DMI) \cite{Roessler2006,2009:Muhlbauer:Science,2010:Yu:Nature, 2012:Seki:Science}. Its phase diagram includes helical (H), conical (C) and skyrmion lattice (SkL) phases. Each phase is known to support characteristic spin excitations, introduced as modes +Q and $-$Q in phases H and C, as well as clockwise (CW), counterclockwise (CCW) and breathing (BR) modes in the SkL phase~\cite{2012:Mochizuki:PhysRevLett,2012:Onose:PhysRevLett,Schwarze2015}. We observe the linearly polarized magnetization dynamics for the modes +Q and $-$Q of the spin-helix phase. Their frequencies are near 2 GHz which are three orders of magnitude smaller compared to easy-plane antiferromagnets \cite{Sievers::PhysRev::1963}. Surprisingly, the linear polarization is present also in almost circularly shaped samples. Furthermore, the polarization characteristics are modified and controlled by fields on the order of 10$^{-2}$ T. This is not possible with antiferromagnets. A ferrimagnet with DMI thus opens an unexpected perspective for the polarization control of microwaves in frequency regimes that are key for modern telecommunication networks. Our findings are paradigm-shifting and a breakthrough for the efficient integration of magnets into future microwave circuitry. Magnetic insulators such as \cso\ are particularly interesting as they provide negligible eddy current loss and small spin-wave damping, which adds to their fascinating topological properties~\cite{2009:Neubauer:PhysRevLett,2010:Jonietz:Science,2012:Schulz:NaturePhys,2013:Milde:Science,2013:Fert:NatureNano,2013:Nagaosa:NatureNano}.

\section*{Results}
\subsection*{Broadband spectroscopy across magnetic phase diagram}
\begin{figure}[t!]
\includegraphics[width=14cm]{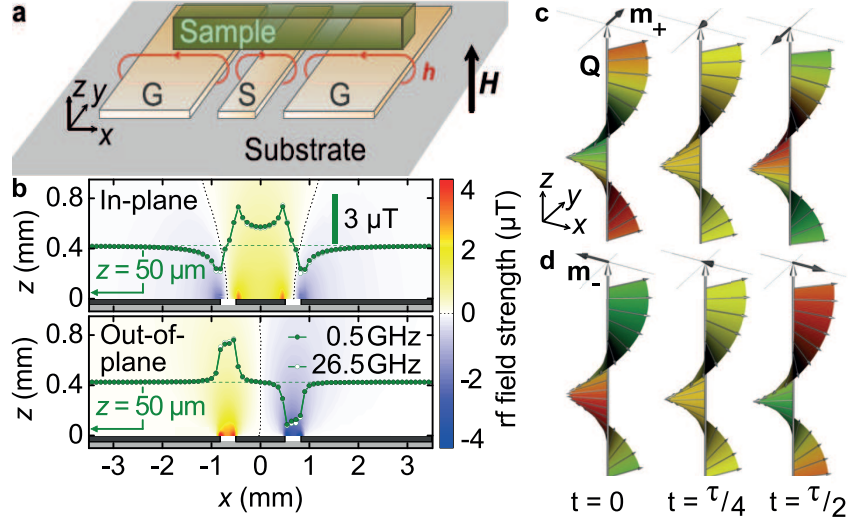}
\caption{\textbf{Coplanar waveguide field distribution and spin precessional motion in the magnetic helix. a},~Sketch of a bar-shaped sample on top of a CPW. \textbf{b},~Color-coded field components $h_{x}$~(top) and $h_{z}$~(bottom) modelled by finite-element electromagnetic simulations~\cite{CST} for a CPW with a 1~mm wide signal line. Green data points illustrate $h_{x}(x)$ and $h_{z}(x)$ at a height of $z = 50\,\mu$m exhibiting both a maximum strength of about 3 $\mu$T. A power of 1\,mW and an impedance of $50\,\Omega$ were considered. \textbf{c},\textbf{d}, Illustrations of precessional motion of spins (thin gray arrows) and averaged dynamic magnetization $\mathbf{m}_{\sigma}(x,y,t)$ with $\sigma=+,-$ (dark arrows) for selected times $t$ during a period $\tau$ of modes (\textbf{c}) +Q and (\textbf{d}) $-$Q. Colors indicate the phase evolution along the helical propagation vector $\mathbf{Q}$. Red (green) color highlights smaller (larger) misalignment angle between neighboring spins compared to their equilibrium position (yellow). We modelled our sample with $N_x<N_y$ at $H=0$, inducing the extraordinary linear polarization of $\mathbf{m}$ (Supplementary Videos 1-4).}
\label{figure1}
\end{figure}

For our experiments we mounted a bar-shaped single crystal of \cso\ with lateral dimensions $2.3\times0.4\times0.3\,\mathrm{mm}^{3}$ on a CPW with a 1 mm-wide signal line [Fig.~\ref{figure1}]. The assembly was cooled down to a temperature $T<T_{\rm c}$, i.e., $T$ was smaller than the critical temperature $T_{\rm c}$ below which the spin-helix phase is stabilized. A static magnetic field $\mathbf{H}$ was applied perpendicular to the substrate, i.e.~along the $z$ axis, being collinear with a $\langle100\rangle$ axis of the crystal. Considering a placement as shown in Fig.~\ref{figure1} and assuming $N_{x} + N_{y} + N_{z} = 1$ for the approximated ellipsoid with $\mathbf{H}$ parallel to a semi-principal axis, we estimated the components of the sample's demagnetization tensor to have components $N_{x} = 0.07$, $N_{y} = 0.40$, and $N_{z} = 0.53$~\cite{Aharoni1998}. Note that $N_{x}$ and $N_{y}$ {\em interchanged} their values if the sample was {\em rotated by $90^{\circ}$} in the $xy$ plane~\cite{Gurevich}. Radiofrequency (rf) signals were applied to the CPW using a vector-network analyzer. The dynamic field $\mathbf{h}$ created by the linearly-polarized electromagnetic wave in the CPW is depicted in Fig.~\ref{figure1}b. We find a dominant in-plane component $h_{x}$ above the signal line and a pronounced out-of-plane component $h_{z}$ above the gaps between the signal (S) and ground (G) lines (for additional experimental details see Methods, Supplementary Notes I and Supplementary Figure 1).

To compare with previous reports on the GHz magnetization dynamics of \cso\ \cite{2012:Onose:PhysRevLett,Schwarze2015} we show different sets of spectra taken throughout the magnetic phase diagram at $T=57~$K$<T_{\rm c}$. In particular we present spectra taken for different sample orientations (Fig.~\ref{figure2}).  In Fig.~\ref{figure2}a we place the 0.4 mm wide sample on the central axis of the 1 mm wide signal line of the CPW similar to earlier setups \cite{2012:Onose:PhysRevLett}. Here, the component $h_x$ of $\mathbf{h}$ provides the relevant torque for spin excitation~\cite{Gurevich}. At $H=0$, i.e.~in the helical phase, we resolve a single resonance at $f\approx1.8$ GHz. For increased $H$, the sample enters the conical phase and this resonance shifts to a smaller frequency (detailed spectra are found in the Supplementary Figure 2). At intermediate fields in Fig.~\ref{figure2}a, two weak resonances are observed (red lines at 18 and 24 mT). We attribute the corresponding field regime to the SkL phase \cite{2012:Onose:PhysRevLett}. Following Refs.~\cite{2012:Mochizuki:PhysRevLett,2012:Onose:PhysRevLett}, $h_x$ excites the modes CCW and CW. The allocation of magnetic phases is consistent with both Ref.\,[\onlinecite{2012:Onose:PhysRevLett}] and measurements on thermodynamic properties performed on \cso\ (such as specific heat or magnetic susceptibility; not shown).
For the sample being collinear with the CPW and within the gap between ground and signal line [Fig.~\ref{figure2}b], we observe a similar sequence of spectra as a function of $H$. A detailed analysis [Fig.~\ref{figure2}e] of eigenfrequencies $f$ in the SkL phase reveals, however, a different field dependence of the prominent SkL mode residing at a higher frequency compared to the setup in Fig.\,\ref{figure2}a. We relate this difference to the out-of-plane component $h_z$ of $\mathbf{h}$ [Fig.~\ref{figure1}b] favoring the excitation of a prominent breathing mode BR~\cite{2012:Mochizuki:PhysRevLett,2012:Onose:PhysRevLett}.\\ \indent When rotating the sample by 90$^\circ$ [Fig.~\ref{figure2}c], spectra at finite field significantly change. At 22 mT, in the conical state, we resolve two modes that are of similar signal strength. With increasing $H$, the low-frequency mode successively vanishes (compare spectra taken at 38 and 46 mT). In the SkL phase, all three modes CCW, BR and CW are seen in one-and-the-same spectrum (compare 28 mT and Supplementary Figure 2), due to simultaneous excitation via both $h_x$ (signal line) and $h_z$ (gap). Still, only one resonance is seen at $H=0$ (helical phase) in Fig.~\ref{figure2}c. Strikingly, for a rotation of 45$^\circ$ (Fig.~\ref{figure2}d) we identify {\em two} modes at $H=0$. The diagonal positioning was not reported before. In the following, we explain the observation of the two modes by an extraordinary linear polarization of magnetization dynamics in the helical phase.

\subsection*{Comparison of experiment with theory}
\begin{figure}[t!]
	\includegraphics[width=14cm]{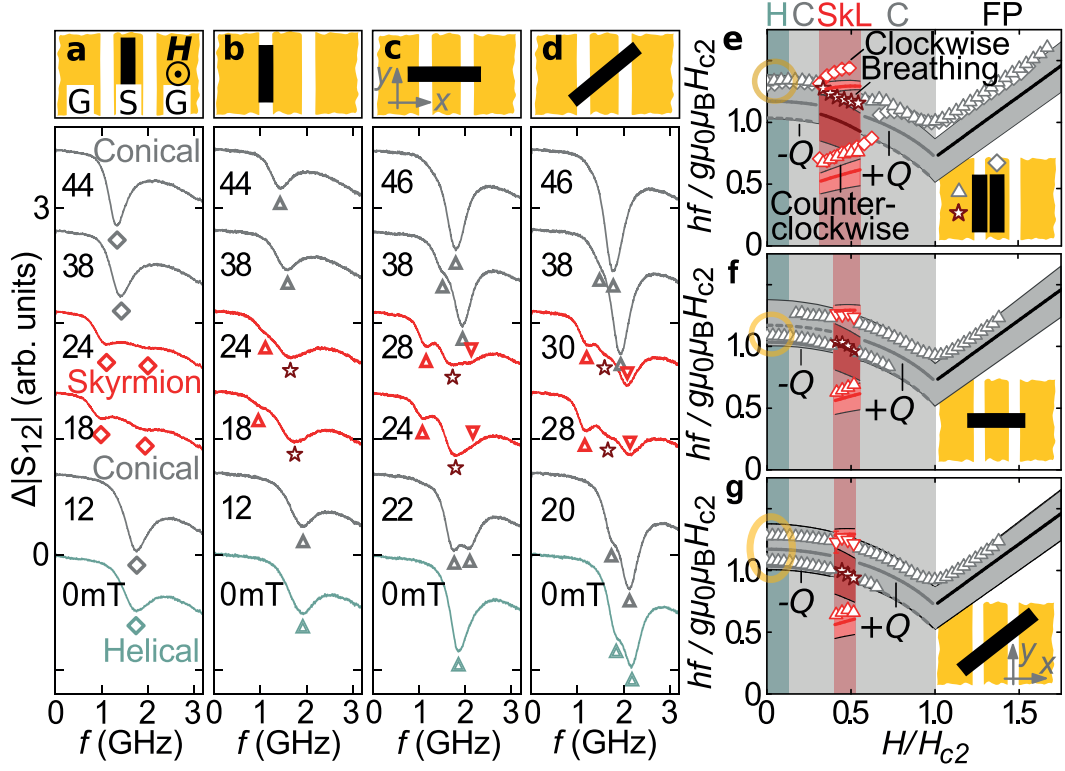}
	\caption{\textbf{Spin resonance data and comparison with theory. \mbox{a--d}},~Typical spectra $\Delta |S_{\text{12}}|$ in arbitrary units with the sample positioned differently on the CPW as sketched on top of each panel. Data are shown for different applied field values in the helical, conical, and SkL phase and are offset for clarity. The field is applied along the $\langle100\rangle$ direction. Symbols indicate resonance frequencies. \textbf{\mbox{e--g}},~Comparison of measured and calculated resonance frequencies. Lines and shaded bands correspond to calculations with $k = 0$ and magnetostatic waves with small $k \neq 0$, respectively~\cite{Schwarze2015}. Dashed lines indicate modes with small spectral weight. Circles highlight that a different number of modes is resolved at small $H$.}
	\label{figure2}
\end{figure}

To compare to theory, it is instructive to summarize both the different number of modes and measured eigenfrequencies $f$ in normalized units following Ref.~\cite{Schwarze2015} in Fig.~\ref{figure2}e-g. The magnetic phases H, C, SkL and FP (field-polarized phase) are indicated by different background colors. From Fig.~\ref{figure2}e-g we find that, qualitatively, in all the different magnetic phases, the measured eigenfrequencies (symbols) follow well the field dependencies (lines) predicted by the theory of Ref.\,[\onlinecite{Schwarze2015}] if we assume a finite wave vector $k$ transferred by the CPW. A nonzero wave vector is provided by the inhomogeneity of $\mathbf{h}(x,z)$ [Fig.\,\ref{figure1}b]. The wave-vector dependence will be discussed elsewhere.

In the following, we focus on the different numbers of modes that we resolved experimentally at small $H$ for phases H and C. Figure~\ref{figure2}g displays that for the sample placed at 45$^\circ$ with respect to the CPW the two resonances detected at $H=0$ are consistent with the two helical modes +Q and $-$Q. This diagonal placement used in Fig.~\ref{figure2}g is usually not executed in the literature. In Figs.~\ref{figure2}e and \ref{figure2}f, where we used the conventional placement, only a single resonance is detected at $H=0$. In Fig.~\ref{figure2}e, theory attributes the single resonance in phases H and C to mode +Q. We do not detect mode $-$Q when the long axis of the sample and the CPW are collinear. Changing the sample orientation by 90$^\circ$ on the CPW (Fig.~\ref{figure2}f), the single resonance detected near $H=0$ is the complementary mode $-$Q. Mode +Q is no longer resolved. Our experiments suggest that at small $H$ a cross-polarization exists between the linearly polarized excitation field $\mathbf{h}$ and either mode $-$Q or $+$Q in Fig.~\ref{figure2}e and \ref{figure2}f, respectively. The corresponding linear polarization of spin helix modes +Q and $-$Q occurring near 2 GHz has not yet been considered. Note that the detection of the ferromagnetic resonance in the FP phase does not depend on the sample placement [Fig.~\ref{figure2}e-g]. The DMI-induced spin texture at low $H$ is thus decisive for the extraordinary polarization characteristics. The occurrence or absence of a double peak at $H=0$ was not explained before~\cite{2012:Onose:PhysRevLett}.

\section*{Discussion}
\subsection*{Polarization and ellipticity}
We now explain the observed polarization dependencies by considering the ellipticity of magnetization precession in the spin helix phases (Fig.~\ref{figure3}). For this, we make use of the theoretical approach outlined in detail in the Supplementary Information of Ref.\,[\onlinecite{Schwarze2015}]. There, spectral weights $\Gamma$ of different modes were discussed, but ellipticities and polarization were not evaluated.
An rf field with $\mathbf{h}$ in a plane perpendicular to ${\bf Q}$ excites the two modes +Q and -Q with a weight that depends on its polarization within this plane, as we show in the following. It is instructive to introduce the corresponding homogeneous dynamic magnetization $\mathbf{m}_{\sigma}$ averaged along a helix period ($\sigma = +1$ and $-1$ for the +Q and $-$Q mode, respectively) [Fig.~\ref{figure1}c,d]. For a helix with $\mathbf{Q}$ in $z$ direction, $\mathbf{m}_{\sigma}=(m^{x}_{\sigma},m^{y}_{\sigma})$ oscillates in the $xy$ plane, characterized by its handedness and ellipticity $|\varepsilon_\sigma| = \left|{m^{x}_{\sigma}}^{2} - {m^{y}_{\sigma}}^{2}\right|^{1/2}/\,\mathrm{max}[m^{x}_{\sigma},m^{y}_{\sigma}]$. The handedness of the ${\bf m}_\sigma$ oscillations is always counterclockwise (left-circular) and clockwise (right-circular) for the +Q and -Q mode, respectively. Here we consider $\mathbf{Q}$ pointing towards the observer.\\ \indent
For pedagogical reasons, we first discuss the polarization of $\mathbf{m}_{\sigma}$ in a disk-shaped sample ($N_x=N_y)$ at $H$=0 (Fig.~\ref{figure3}a). In this case, the oscillation is circular with zero ellipticity for symmetry reasons. As a linearly polarized excitation field $\mathbf{h}$ in the $xy$ plane is equivalent to the superposition of left- and right-circularly polarized components, it couples equally well to the two modes +Q and -Q and leads to the same nonzero spectral weight $\varGamma_\sigma$ for both modes. Note that for a disk with $N_x = N_y$, the screw symmetry of the helix ensures that the two excitation modes are degenerate in frequency at $H=0$. By appropriately combining the two circularly polarized modes, any polarization can be achieved. For $H \neq 0$ applied along $\mathbf{Q}$, the precessional motion of each individual mode stays circularly polarized. The handedness of the +Q mode (left-circular) coincides with the one of the local spin precession and thus with the Kittel mode in the FP phase. Mode +Q smoothly connects to the Kittel mode at $H_{c2}$. Contrastingly, the weight of the -Q mode reduces to zero close to the critical field $H_{c2}$. This is because mode $-$Q is right-circular and does not comply to the handedness imposed by the Landau-Lifshitz equation of motion~\cite{Gurevich}. The resulting imbalance in precession amplitude compared to +Q is indicated by bold and broken lines in Fig.~\ref{figure3}c (see $\varGamma_\sigma$ at $N_x=0.235$). The imbalance increases with increasing $H$. For a disk-shaped sample with $N_x = N_y$ (=~0.235 in our case) the modes $\pm Q$ are thus circularly polarized for all fields $H<H_{\rm c2}$ [Fig.\,\ref{figure3}a and Supplementary Videos 5-8 (at $H=0$) and 9-12 (at $H=0.1\,H_{\rm c2}$)] and exhibit $\varepsilon_\sigma = 0$.

For a sample with ellipsoidal shape within the $xy$ plane, i.e., $N_x \neq N_y$, the polarization of the $\pm$ Q modes is no longer circular and $\varepsilon_\sigma$ becomes nonzero. This is illustrated in the main panel of Fig.~\ref{figure3}b, where we show ellipticities calculated for $0 \leq N_{x} \leq 1 - N_{z}$ at different field values. We set $\varepsilon_{\sigma}$ positive and negative for the long principal axis being along $\hat{x}$ and $\hat{y}$, respectively [sketches on the right side of Fig.~\ref{figure3}b].

We first focus on $\varepsilon_{\sigma}$ in zero field (black lines). As soon as $N_x \neq N_y$ the frequency degeneracy of modes $+$Q and $-$Q is lifted, and the modes are found to be strictly linearly polarized. In zero field, the spins of the chiral magnet align perpendicular to the helical propagation vector \textbf{Q}, so that the helix possesses a 180$^{\circ}$-rotation symmetry around each of its spins. By virtue of this symmetry, there exist always pairs of spins within the helix whose precessional motion conspire to yield a linear polarization along one of the principal axes, i.e.~$x$ or $y$. This scenario is similar to the easy-plane AFM \cite{Sievers::PhysRev::1963}. As a consequence, the ellipticities of modes +Q (solid line) and $-$Q (dotted line) follow step functions $\varepsilon_\sigma|_{H=0} = \mathrm{sgn}[\sigma\cdot(N_{x} - N_{y})]$ exhibiting $\sigma = \pm 1$ with an unexpectedly pronounced sensitivity regarding the sample shape. This striking sensitivity is known neither from electronic excitations nor from ferrimagnets or easy-plane antiferromagnets.
\subsection*{Spectral weights}
\begin{figure}[t!]
	\includegraphics[width=12cm]{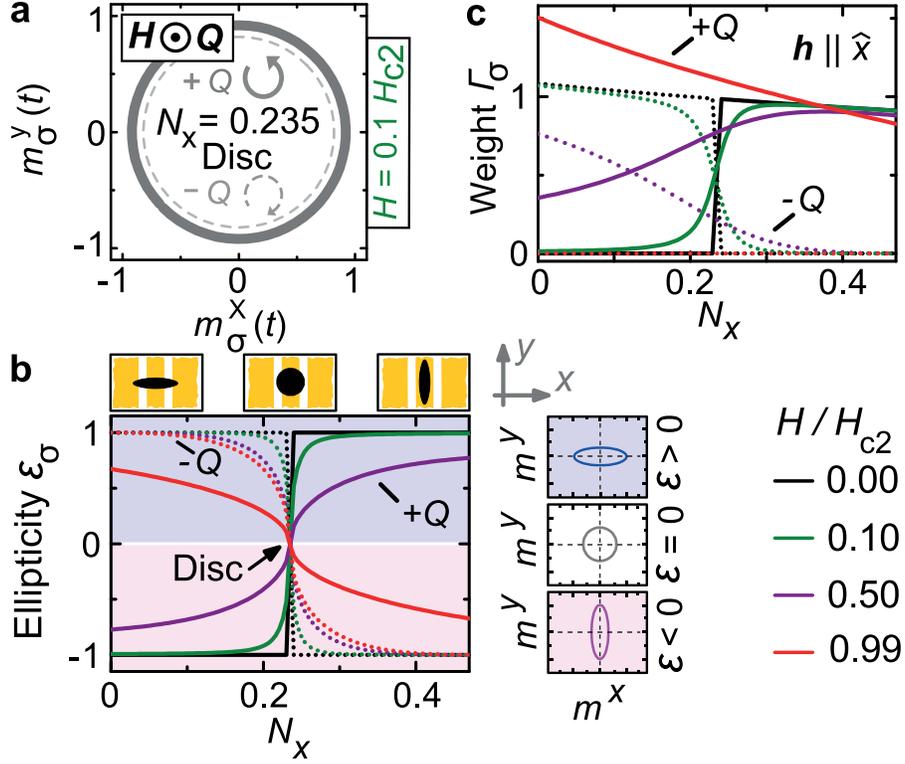}
	\caption{\textbf{Ellipticity and spectral weight. a},~Dynamic components of the circularly polarized modes +Q and $-$Q in a round, flat sample ($N_{x} = N_{y} < N_{z}$) with a small field $H$ applied along $\hat{z}$. The width of lines indicates the relative signal strength. \textbf{b},~Ellipticity $\varepsilon_\sigma$ of the conical modes as a function of the demagnetization factor $N_{x}$ ($N_{z} = 0.53$) calculated for different magnetic fields~(colors). Solid and dotted lines represent modes +Q and $-$Q. The insets illustrate the sample shapes corresponding to $N_{x}$~(top) and the elliptic trajectories of the average spin precessional motion~(right). \textbf{c},~Spectral weight $\varGamma_{\sigma}$ of the conical modes (arb.~units) for an excitation field $\mathbf{h}$ along $\hat{x}$.}
	\label{figure3}
\end{figure}

The characteristic shape dependence of $\varepsilon_\sigma$ of a DMI-containing magnet has an impact on the spectral weights $\varGamma_{\sigma}$ [Fig.~\ref{figure3}c]. For the spectral weight of the $\pm Q$ modes at $H=0$ we find
\begin{equation}
\varGamma_{\sigma}(N_{x},N_{y})\big|_{H=0} \propto \frac{\varTheta[\sigma\cdot(N_{x} - N_{y})]}{1+(2+N_{x})\chi_{\mathrm{con}}^{\mathrm{int}}/6}
\end{equation}
with the Heaviside step function $\varTheta(s) = 1$ for $s > 0$ and zero otherwise ($\chi_{\mathrm{con}}^{\mathrm{int}}$ is the internal conical susceptibility \cite{Schwarze2015}). Here we consider an rf field $\bf h$ along the $x$ direction. Hence, selective excitation of either mode +Q or $-$Q can be realized. For instance, mode +Q is polarized along $\hat{y}$ ($\epsilon_+= -1$) for $N_{x} < N_{y}$ [Fig.~\ref{figure1}c]. An rf field $\mathbf{h}$ along $\hat{x}$ does not couple to its precessional motion and, as a consequence, the spectral weight $\varGamma_{+}$ is zero in the helical phase at $H=0$ [full black line in Fig.~\ref{figure3}c for $N_{x} <0.235$]. For mode $-$Q, $\epsilon_- = +1$ [polarized along $\hat{x}\,||\,\mathbf{h}$ in Fig.~\ref{figure1}d] and $\varGamma_{-}$ is large [dotted black line in Fig.~\ref{figure3}c for $N_{x} <0.235$].
\\
\indent
For $H > 0$, spins within the helix cant towards the field direction forming the conical state. Note that considering the ellipticities and spectral weights of both modes in Fig.~\ref{figure3}b and c, respectively, their field dependencies are not symmetric with respect to $N_{x} =0.235$ (disk-like shape). For $N_{x} \ll N_{y}$, $\varepsilon_{-}$ of mode $-$Q stays close to $+1$ [colored dotted lines in Fig.~\ref{figure3}b for $N_{x} <0.235$ and Supplementary Videos 13-16] and its linear polarization persists over a remarkably large field range. We extract $\varepsilon_{-}=0.999$ at $H = 0.5\,H_{c2}$ for $N_{x} = 0.07$ (our sample) so that an rf field ${\bf h} \parallel \hat x$ can couple to the -Q mode. Nevertheless, the spectral weight $\Gamma_-$ decreases significantly with increasing $H$ towards $H_{c2}$ (colored dotted lines in Fig.\,\ref{figure3}c) because of the mismatch of handedness with the Kittel mode within the FP phase as discussed above.
Conversely, $\varGamma_{+}$ grows much with $H$ as the ellipticity $|\varepsilon_{+}|$ of the mode +Q first decreases reaching circular polarization at $H_{\rm circ}\approx 0.76\,H_{\rm c2}$ and then increases again staying below $\varepsilon_{+}=1$ [colored full lines in Fig.~\ref{figure3}b, Supplementary Equations (5)-(8) and Supplementary Notes II]. These features explain our experimental observations in Fig.~\ref{figure2}c and \ref{figure2}f in which signal strengths of mode $-$Q and +Q interchange completely with increasing $H$. The diagonal placement of Fig.~\ref{figure2}d allows us to detect both modes +Q and $-$Q at $H=0$ as $\mathbf{h}\,||\,\hat{x}$ can be decomposed into equal components along and transverse to the bar-shaped \cso. Still, the signal strengths of mode +Q and $-$Q are found to be different at $H=0$. We have applied the same formalism to the three eigenmodes of the SkL phase. As the SkL phase only exists at a finite magnetic field $H$, the clockwise and counterclockwise modes are always elliptically polarized with a finite $\varepsilon$, whereas the breathing mode is linearly polarized by symmetry (Supplementary Notes II).

\begin{figure}[t!]
	\includegraphics[width=7cm]{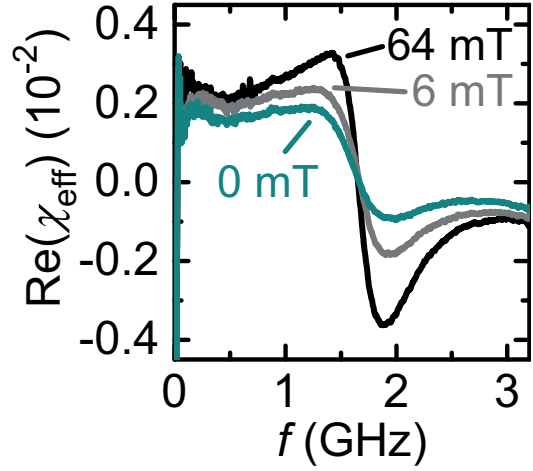}
	\caption{\textbf{Dynamic susceptibility of modes with different polarization and similar eigenfrequency.} Real part of the effective dynamic susceptibility in the helical ($0$\,mT), conical ($6$\,mT), and field-polarised ($64$\,mT) phase at $57.3$\,K. Each magnetic phase provides a different polarization of the magnetization dynamics [see Fig.\,\ref{figure3} (b)]. The signal strength, i.e., peak-to-peak variation of the effective susceptibility, amounts to about 0.007 (0.003) in the field-polarized (helical) phase.}	
	\label{ReS12}
\end{figure}

In Fig.\,\ref{ReS12} we compare the real part of the complex effective microwave susceptibility $\chi_{\rm eff}(f)$ \cite{Kalarickal::JAPL::2006} recorded at $57.3$\,K in different phases of \cso. The resonance positions (inflection points) of the three curves coincide but the signal strength (peak-to-peak amplitude) varies. The largest signal strength is obtained for the field-polarized phase when the magnetization $M$ of the ferrimagnet is maximum. Note that \cso\ and the technologically relevant ferrimagnet YIG exhibit a similar magnetization $\mu_0M$ of $0.13$\,T\,\cite{2012:Adams:PhysRevLett,Schwarze2015} and $0.18$\,T\,\cite{2013:AllivyKelly:APL}, respectively. At $H=0$ where the overall magnetization of the helical phase amounts to zero, we do not find the susceptibility of \cso\ to be reduced significantly compared to its field-polarized phase (Fig.\,\ref{ReS12}). We hence expect an optimized chiral insulator with low damping at room temperature to become technologically relevant. The linear polarization of the helical phase and the related linear dichroism in the few GHz frequency regime open new avenues for microwave components. For example, a conventional linear polarizer for free-space GHz radiation consists of a large grid prepared from macroscopic metallic rods mounted on a large frame. Contrastingly, our results offer a field-tunable magnetic polarizer based on a compact material. The linear polarization in chiral magnets reported in our manuscript is very robust and arises for any generic sample shape deviating from a circular shape, as it is controlled by the symmetries of the helical spin structure that is stabilized by the Dzyaloshinskii-Moriya-interaction. Magnon-photon cavities is one of many future technologies where  DMI-induced linear dichroism offers novel, exciting perspectives for microwave control with integrated magnetic components by boosting the coupling between linearly polarized electromagnetic waves and magnons.

In conclusion, we studied collective spin excitations of an insulating ferrimagnet coupled to a GHz magnetic field and discovered the shape-controlled linear polarization of cross-polarized spin-helix modes. In our experiments we selected the complementary spin-helix modes by rotation of the sample, i.e., by interchanging demagnetization factors. To explain this, we explored the ellipticity of magnetization dynamics in the presence of Dzyaloshinskii-Moriya interaction. On the one hand, the extraordinary linear polarization of magnetization precession is of fundamental interest as it is not known from the extensive studies on other magnetic materials at low GHz frequencies. On the other hand, it allows for the efficient exploitation of magnetization dynamics by standard microwave equipment such as cavities or coplanar waveguides offering a high degree of on-chip integration.

\section*{Methods}
The crystals were grown by means of chemical vapor transport using HCl as transport agent \cite{Gnezdilov2010} and  crystallized as cubic chiral magnets in space group $P2_{1}3$. \cso\ is considered as a local-moment ferrimagnetic insulator~\cite{1977:Kohn:JPhysSocJpn, 2010:Belesi:PhysRevB,Janson2014}. The combination of exchange and Dzyaloshinskii-Moriya interaction (DMI) results in long-wavelength helimagnetism (wavelength $\lambda_{\mathrm{h}} = 620\,\mathrm{\AA}$) below a critical field $H_{c2}$ and below $T_{c}$~\cite{2012:Adams:PhysRevLett, 2012:Seki:PhysRevB}.
We connected a vector network analyzer to both ends of the CPW to excite and detect the spin dynamics by measuring the scattering parameter $S_{12}$ in transmission configuration \cite{Schwarze2015}. A reference spectrum was subtracted to obtain $\Delta|S_{12}|$ and enhance the signal-to-noise ratio. We do not address the nonreciprocal directional dichroism attributed to magnetoelectric coupling in \cso\ ~\cite{2012:Seki:PhysRevB2,2012:White:JPhysCondensMatter,2013:Okamura:NatCommun,WhitePRL2014, 2015::Okamura, Mochizuki2015PRL} and therefore limit our discussion to the magnitude of the parameter $|S_{12}|$. Note that for a substrate with a relative dielectric constant $\epsilon_{\rm r}=1$, electrical and magnetic components of the electromagnetic wave are linearly polarized in a 50 $\Omega$ matched CPW~\cite{Wen1969}. We assumed our Rogers substrate to approximate this condition as $\epsilon_{\rm r}=3.5$ was small. The spatial distribution of $\mathbf{h}$ does not vary much in the few GHz frequency regime addressed here. The field dependency of eigenfrequencies $f$, ellipticity $\epsilon$ and spectral weight $w$ of modes were modelled using the theoretical approach of Ref.~[\onlinecite{Schwarze2015}] that included, both, DMI and dipolar effect due to the sample shape. Note that the propagation vector $\mathbf{Q}$ for helical and conical modes +Q and $-$Q [Fig.~\ref{figure1}c,d] was oriented along the $z$ axis ($\hat{z}$) in our experiments as we measured spectra after saturating the sample along the corresponding $\langle100\rangle$ axis being parallel to $\hat{z}$.
For the analysis in Fig.~\ref{figure2}e-g, the frequencies and field values are normalized as introduced in Ref. [\onlinecite{Schwarze2015}]. The normalization allows us to present consistently the different datasets taking into account the slight variation of temperature $T$ when remounting the sample in a different orientation (Supplementary Notes I). The lines indicate eigenfrequencies for different modes as predicted by the theoretical approach outlined in Ref.\,[\onlinecite{Schwarze2015}]. When extracting the susceptibility from scattering parameters $S_{\rm 12}$ we corrected for the slight impedance mismatch between the sample and CPW by adjusting the delay time of the electromagnetic wave. The real part of the complex effective microwave susceptibility $\chi_{\rm eff}(f)$ \cite{Kalarickal::JAPL::2006}
is calculated via $\chi_{\rm eff}(f)=i\,\Bigg(\frac{\text{ln}\big[S_{\rm 12}^\text{meas}(f)\big]}{\text{ln}\big[S_{\rm 12}^\text{ref}(f)\big]}-1\Bigg)$, where $f$ is the frequency, $S_{\rm 12}^\text{meas}$ is the scattering parameter measured in the corresponding magnetic phase and $S_{\rm 12}^\text{ref}$ represents a reference spectrum containing the field-independent background signal.

\section*{Acknowledgements}
We thank S.\ Mayr for assistance with sample preparation, as well as A.\ N.\ Slavin, M.\ Bailleul, N.\ Kanazawa, F.\ Lisiecki, and S.\ Watanabe for fruitful discussions and support on the design of antennas for broadband spectroscopy. Financial support through DFG TRR80 (From Electronic Correlations to Functionality), DFG FOR960 (Quantum Phase Transitions), and ERC AdG (291079, TOPFIT) is gratefully acknowledged. A.B.\ acknowledges financial support through the TUM graduate school.

\section*{Author contributions}
I.S., J.W. and D.G. planned the experiment. I.S. and S.W. conducted the experiments. I.S., S.W., A.B., D.G. analysed the experimental data. H.B. fabricated the samples. J.W. and M.G. performed the theoretical calculations.
D.G. and C.P. proposed this study. I.S, A.B., M.G. and D.G. wrote the manuscript. All authors discussed the data and commented on the manuscript.

\section*{Additional information}
Supplementary information is available in the online version of the paper. Reprints and permissions information is available online at www.nature.com/reprints. Correspondence and requests for materials should be addressed D.G.

\section*{Competing financial interests}
The authors declare no competing financial interests.

\begin{thebibliography}{10}
	\expandafter\ifx\csname url\endcsname\relax
	\def\url#1{\texttt{#1}}\fi
	\expandafter\ifx\csname urlprefix\endcsname\relax\def\urlprefix{URL }\fi
	\providecommand{\bibinfo}[2]{#2}
	\providecommand{\eprint}[2][]{\url{#2}}
	
	\bibitem{LanLif1935}
	\bibinfo{author}{Landau, L.} \& \bibinfo{author}{Lifshitz, E.}
	\newblock \bibinfo{title}{On the theory of the dispersion of magnetic
		permeability in ferromagnetic bodies}.
	\newblock \emph{\bibinfo{journal}{Physikalische Zeitschrift der Sowjetunion}}
	\textbf{\bibinfo{volume}{8}}, \bibinfo{pages}{153 -- 169}
	(\bibinfo{year}{1935}).
	
	\bibitem{Gurevich}
	\bibinfo{author}{Gurevich, A.~G.} \& \bibinfo{author}{Melkov, G.~A.}
	\newblock \emph{\bibinfo{title}{Magnetization Oscillations and Waves}}
	(\bibinfo{publisher}{CRC Press}, \bibinfo{address}{Boca Raton},
	\bibinfo{year}{1996}).
	
	\bibitem{Gruenberg::NobelLecture}
	\bibinfo{author}{Gr\"unberg, P.~A.}
	\newblock \bibinfo{title}{Nobel lecture: From spin waves to giant
		magnetoresistance and beyond}.
	\newblock \emph{\bibinfo{journal}{Rev. Mod. Phys.}}
	\textbf{\bibinfo{volume}{80}}, \bibinfo{pages}{1531--1540}
	(\bibinfo{year}{2008}).
	\newblock \urlprefix\url{http://link.aps.org/doi/10.1103/RevModPhys.80.1531}.
	
\bibitem{Zivieri2002}
\bibinfo{author}{Zivieri, R.} \emph{et~al.}
\newblock \bibinfo{title}{Stokes\char21{}anti-stokes brillouin intensity
  asymmetry of spin-wave modes in ferromagnetic films and multilayers}.
\newblock \emph{\bibinfo{journal}{Phys. Rev. B}} \textbf{\bibinfo{volume}{65}},
  \bibinfo{pages}{165406} (\bibinfo{year}{2002}).
\newblock \urlprefix\url{http://link.aps.org/doi/10.1103/PhysRevB.65.165406}.

\bibitem{Klausen2004}
\bibinfo{author}{Klausen, S.~N.} \emph{et~al.}
\newblock \bibinfo{title}{Magnetic anisotropy and quantized spin waves in
  hematite nanoparticles}.
\newblock \emph{\bibinfo{journal}{Phys. Rev. B}} \textbf{\bibinfo{volume}{70}},
  \bibinfo{pages}{214411} (\bibinfo{year}{2004}).
\newblock \urlprefix\url{http://link.aps.org/doi/10.1103/PhysRevB.70.214411}.
	\bibitem{Sievers::PhysRev::1963}
	\bibinfo{author}{Sievers, A.~J.} \& \bibinfo{author}{Tinkham, M.}
	\newblock \bibinfo{title}{Far infrared antiferromagnetic resonance in {M}n{O}
		and {N}i{O}}.
	\newblock \emph{\bibinfo{journal}{Phys. Rev.}} \textbf{\bibinfo{volume}{129}},
	\bibinfo{pages}{1566--1571} (\bibinfo{year}{1963}).
	\newblock \urlprefix\url{http://link.aps.org/doi/10.1103/PhysRev.129.1566}.
	
	\bibitem{Kampfrath08}
	\bibinfo{author}{Kampfrath, T.} \emph{et~al.}
	\newblock \bibinfo{title}{Coherent terahertz control of antiferromagnetic spin
		waves}.
	\newblock \emph{\bibinfo{journal}{Nat. Photon.}} \textbf{\bibinfo{volume}{5}},
	\bibinfo{pages}{31--34} (\bibinfo{year}{2011}).
	\newblock \urlprefix\url{http://dx.doi.org/10.1038/nphoton.2010.259}.

	\bibitem{Huebl2013}
	\bibinfo{author}{Huebl, H.} \emph{et~al.}
	\newblock \bibinfo{title}{High cooperativity in coupled microwave resonator
		ferrimagnetic insulator hybrids}.
	\newblock \emph{\bibinfo{journal}{Phys. Rev. Lett.}}
	\textbf{\bibinfo{volume}{111}}, \bibinfo{pages}{127003}
	(\bibinfo{year}{2013}).
	\newblock
	\urlprefix\url{http://link.aps.org/doi/10.1103/PhysRevLett.111.127003}.
	
	\bibitem{Tabuchi2014}
	\bibinfo{author}{Tabuchi, Y.} \emph{et~al.}
	\newblock \bibinfo{title}{Hybridizing ferromagnetic magnons and microwave
		photons in the quantum limit}.
	\newblock \emph{\bibinfo{journal}{Phys. Rev. Lett.}}
	\textbf{\bibinfo{volume}{113}}, \bibinfo{pages}{083603}
	(\bibinfo{year}{2014}).
	\newblock
	\urlprefix\url{http://link.aps.org/doi/10.1103/PhysRevLett.113.083603}.
	
	\bibitem{Zhang2014}
	\bibinfo{author}{Zhang, X.}, \bibinfo{author}{Zou, C.-L.},
	\bibinfo{author}{Jiang, L.} \& \bibinfo{author}{Tang, H.~X.}
	\newblock \bibinfo{title}{Strongly coupled magnons and cavity microwave
		photons}.
	\newblock \emph{\bibinfo{journal}{Phys. Rev. Lett.}}
	\textbf{\bibinfo{volume}{113}}, \bibinfo{pages}{156401}
	(\bibinfo{year}{2014}).
	\newblock
	\urlprefix\url{http://link.aps.org/doi/10.1103/PhysRevLett.113.156401}.
	
	\bibitem{Goryachev2014}
	\bibinfo{author}{Goryachev, M.} \emph{et~al.}
	\newblock \bibinfo{title}{High-cooperativity cavity qed with magnons at
		microwave frequencies}.
	\newblock \emph{\bibinfo{journal}{Phys. Rev. Applied}}
	\textbf{\bibinfo{volume}{2}}, \bibinfo{pages}{054002} (\bibinfo{year}{2014}).
	\newblock
	\urlprefix\url{http://link.aps.org/doi/10.1103/PhysRevApplied.2.054002}.
	
	\bibitem{Wen1969}
	\bibinfo{author}{Wen, C.~P.}
	\newblock \bibinfo{title}{Coplanar waveguide: a surface strip transmission line
		suitable for nonreciprocal gyromagnetic device applications}.
	\newblock \emph{\bibinfo{journal}{IEEE Trans. Microwave Theory and Techniques}}
	\textbf{\bibinfo{volume}{MIT-17}}, \bibinfo{pages}{1087--1090}
	(\bibinfo{year}{1969}).
	
	\bibitem{Duncan1957}
	\bibinfo{author}{Duncan, B.~J.}, \bibinfo{author}{Swern, L.},
	\bibinfo{author}{Tomiyasu, K.} \& \bibinfo{author}{Hannwacker, J.}
	\newblock \bibinfo{title}{Design considerations for broad-band ferrite coaxial
		line isolators}.
	\newblock \emph{\bibinfo{journal}{Proceedings of the IRE}}
	\textbf{\bibinfo{volume}{45}}, \bibinfo{pages}{483--490}
	(\bibinfo{year}{1957}).
	
	\bibitem{Poole67}
	\bibinfo{author}{Poole, C.~P.}
	\newblock \emph{\bibinfo{title}{Electron Spin Resonance \textemdash\ A
			Comprehensive Treatise on Experimental Techniques}}
	(\bibinfo{publisher}{Interscience {P}ublishers}, \bibinfo{address}{New York},
	\bibinfo{year}{1967}).
	
	\bibitem{Yasukawa2016}
	\bibinfo{author}{Yasukawa, T.}, \bibinfo{author}{Sigillito, A.~J.},
	\bibinfo{author}{Rose, B.~C.}, \bibinfo{author}{Tyryshkin, A.~M.} \&
	\bibinfo{author}{Lyon, S.~A.}
	\newblock \bibinfo{title}{Addressing spin transitions on $^{209}\mathrm{Bi}$
		donors in silicon using circularly polarized microwaves}.
	\newblock \emph{\bibinfo{journal}{Phys. Rev. B}} \textbf{\bibinfo{volume}{93}},
	\bibinfo{pages}{121306} (\bibinfo{year}{2016}).
	\newblock \urlprefix\url{http://link.aps.org/doi/10.1103/PhysRevB.93.121306}.

	
	\bibitem{Roessler2006}
	\bibinfo{author}{R{\"o}{\ss}ler, U.~K.}, \bibinfo{author}{Bogdanov, A.~N.} \&
	\bibinfo{author}{Pfleiderer, C.}
	\newblock \bibinfo{title}{Spontaneous skyrmion ground states in magnetic
		metals}.
	\newblock \emph{\bibinfo{journal}{Nature (London)}}
	\textbf{\bibinfo{volume}{442}}, \bibinfo{pages}{797--801}
	(\bibinfo{year}{2006}).
	\newblock \urlprefix\url{http://dx.doi.org/10.1038/nature05056}.
	
	\bibitem{2009:Muhlbauer:Science}
	\bibinfo{author}{M\"{u}hlbauer, S.} \emph{et~al.}
	\newblock \bibinfo{title}{{Skyrmion Lattice in a Chiral Magnet}}.
	\newblock \emph{\bibinfo{journal}{Science}} \textbf{\bibinfo{volume}{323}},
	\bibinfo{pages}{915} (\bibinfo{year}{2009}).
	\newblock \urlprefix\url{http://www.sciencemag.org/content/323/5916/915.full}.
	
	\bibitem{2010:Yu:Nature}
	\bibinfo{author}{Yu, X.~Z.} \emph{et~al.}
	\newblock \bibinfo{title}{{Real-space observation of a two-dimensional skyrmion
			crystal}}.
	\newblock \emph{\bibinfo{journal}{Nature (London)}}
	\textbf{\bibinfo{volume}{465}}, \bibinfo{pages}{901} (\bibinfo{year}{2010}).
	\newblock
	\urlprefix\url{http://www.nature.com/nature/journal/v465/n7300/full/nature09124.html}.
	
	\bibitem{2012:Seki:Science}
	\bibinfo{author}{Seki, S.}, \bibinfo{author}{Yu, X.~Z.},
	\bibinfo{author}{Ishiwata, S.} \& \bibinfo{author}{Tokura, Y.}
	\newblock \bibinfo{title}{{Observation of Skyrmions in a Multiferroic
			Material}}.
	\newblock \emph{\bibinfo{journal}{Science}} \textbf{\bibinfo{volume}{336}},
	\bibinfo{pages}{198} (\bibinfo{year}{2012}).
	\newblock
	\urlprefix\url{http://www.sciencemag.org/content/336/6078/198.full.html}.
	
	\bibitem{2012:Mochizuki:PhysRevLett}
	\bibinfo{author}{Mochizuki, M.}
	\newblock \bibinfo{title}{{Spin-Wave Modes and Their Intense Excitation Effects
			in Skyrmion Crystals}}.
	\newblock \emph{\bibinfo{journal}{Phys. Rev. Lett.}}
	\textbf{\bibinfo{volume}{108}}, \bibinfo{pages}{017601}
	(\bibinfo{year}{2012}).
	\newblock
	\urlprefix\url{http://link.aps.org/doi/10.1103/PhysRevLett.108.017601}.
	
	\bibitem{2012:Onose:PhysRevLett}
	\bibinfo{author}{Onose, Y.}, \bibinfo{author}{Okamura, Y.},
	\bibinfo{author}{Seki, S.}, \bibinfo{author}{Ishiwata, S.} \&
	\bibinfo{author}{Tokura, Y.}
	\newblock \bibinfo{title}{{Observation of Magnetic Excitations of Skyrmion
			Crystal in a Helimagnetic Insulator Cu$_{2}$OSeO$_{3}$}}.
	\newblock \emph{\bibinfo{journal}{Phys. Rev. Lett.}}
	\textbf{\bibinfo{volume}{109}}, \bibinfo{pages}{037603}
	(\bibinfo{year}{2012}).
	\newblock
	\urlprefix\url{http://link.aps.org/doi/10.1103/PhysRevLett.109.037603}.
	
	\bibitem{Schwarze2015}
	\bibinfo{author}{Schwarze, T.} \emph{et~al.}
	\newblock \bibinfo{title}{Universal helimagnon and skyrmion excitations in
		metallic, semiconducting and insulating chiral magnets}.
	\newblock \emph{\bibinfo{journal}{Nature Mater.}}
	\textbf{\bibinfo{volume}{14}}, \bibinfo{pages}{478--483}
	(\bibinfo{year}{2015}).
	\newblock \urlprefix\url{http://dx.doi.org/10.1038/nmat4223}.
	
	\bibitem{2009:Neubauer:PhysRevLett}
	\bibinfo{author}{Neubauer, A.} \emph{et~al.}
	\newblock \bibinfo{title}{{Topological Hall Effect in the $A$ Phase of MnSi}}.
	\newblock \emph{\bibinfo{journal}{Phys. Rev. Lett.}}
	\textbf{\bibinfo{volume}{102}}, \bibinfo{pages}{186602}
	(\bibinfo{year}{2009}).
	\newblock
	\urlprefix\url{http://link.aps.org/doi/10.1103/PhysRevLett.102.186602}.
	
	\bibitem{2010:Jonietz:Science}
	\bibinfo{author}{Jonietz, F.} \emph{et~al.}
	\newblock \bibinfo{title}{{Spin Transfer Torques in MnSi at Ultralow Current
			Densities}}.
	\newblock \emph{\bibinfo{journal}{Science}} \textbf{\bibinfo{volume}{330}},
	\bibinfo{pages}{1648} (\bibinfo{year}{2010}).
	\newblock
	\urlprefix\url{http://www.sciencemag.org/content/330/6011/1648.full.html}.
	
	\bibitem{2012:Schulz:NaturePhys}
	\bibinfo{author}{Schulz, T.} \emph{et~al.}
	\newblock \bibinfo{title}{{Emergent electrodynamics of skyrmions in a chiral
			magnet}}.
	\newblock \emph{\bibinfo{journal}{Nature Phys.}} \textbf{\bibinfo{volume}{8}},
	\bibinfo{pages}{301} (\bibinfo{year}{2012}).
	\newblock
	\urlprefix\url{http://www.nature.com/nphys/journal/v8/n4/full/nphys2231.html}.
	
	\bibitem{2013:Milde:Science}
	\bibinfo{author}{Milde, P.} \emph{et~al.}
	\newblock \bibinfo{title}{{Unwinding of a Skyrmion Lattice by Magnetic
			Monopoles}}.
	\newblock \emph{\bibinfo{journal}{Science}} \textbf{\bibinfo{volume}{340}},
	\bibinfo{pages}{1076} (\bibinfo{year}{2013}).
	\newblock
	\urlprefix\url{http://www.sciencemag.org/content/340/6136/1076.full.html}.
	
	\bibitem{2013:Fert:NatureNano}
	\bibinfo{author}{Fert, A.}, \bibinfo{author}{Cros, V.} \&
	\bibinfo{author}{Sampaio, J.}
	\newblock \bibinfo{title}{{Skyrmions on the track}}.
	\newblock \emph{\bibinfo{journal}{Nature Nano.}} \textbf{\bibinfo{volume}{8}},
	\bibinfo{pages}{152} (\bibinfo{year}{2013}).
	\newblock
	\urlprefix\url{http://www.nature.com/nnano/journal/v8/n3/full/nnano.2013.29.html}.
	
	\bibitem{2013:Nagaosa:NatureNano}
	\bibinfo{author}{Nagaosa, N.} \& \bibinfo{author}{Tokura, Y.}
	\newblock \bibinfo{title}{{Topological properties and dynamics of magnetic
			skyrmions}}.
	\newblock \emph{\bibinfo{journal}{Nature Nano.}} \textbf{\bibinfo{volume}{8}},
	\bibinfo{pages}{899} (\bibinfo{year}{2013}).
	\newblock
	\urlprefix\url{http://www.nature.com/nnano/journal/v8/n12/full/nnano.2013.243.html}.
	
	\bibitem{CST}
	\bibinfo{note}{The software \emph{CST Microwave Studio 2014} (CST Computer
		Simulation Technology, www.cst.com) has been used for optimizing the
		impedance to 50\,$\Omega$ and visualize the field profile of the CPW.}
	
	\bibitem{Aharoni1998}
	\bibinfo{author}{Aharoni, A.}
	\newblock \bibinfo{title}{Demagnetizing factors for rectangular ferromagnetic
		prisms}.
	\newblock \emph{\bibinfo{journal}{J. Appl. Phys.}}
	\textbf{\bibinfo{volume}{83}}, \bibinfo{pages}{3432--3434}
	(\bibinfo{year}{1998}).
	\newblock
	\urlprefix\url{http://scitation.aip.org/content/aip/journal/jap/83/6/10.1063/1.367113}.
	
	\bibitem{Gnezdilov2010}
	\bibinfo{author}{Gnezdilov, V.~P.} \emph{et~al.}
	\newblock \bibinfo{title}{{Magnetoelectricity in the ferrimagnetic
			Cu$_{2}$OSeO$_{3}$: symmetry analysis and Raman scattering study}}.
	\newblock \emph{\bibinfo{journal}{Low Temp. Phys.}}
	\textbf{\bibinfo{volume}{36}}, \bibinfo{pages}{550--557}
	(\bibinfo{year}{2010}).
	\newblock
	\urlprefix\url{http://scitation.aip.org/content/aip/journal/ltp/36/6/10.1063/1.3455808}.
	
	\bibitem{1977:Kohn:JPhysSocJpn}
	\bibinfo{author}{Kohn, K.}
	\newblock \bibinfo{title}{{A New Ferrimagnet Cu$_{2}$SeO$_{4}$}}.
	\newblock \emph{\bibinfo{journal}{J. Phys. Soc. Jpn}}
	\textbf{\bibinfo{volume}{42}}, \bibinfo{pages}{2065} (\bibinfo{year}{1977}).
	\newblock \urlprefix\url{http://jpsj.ipap.jp/link?JPSJ/42/2065/}.
	
	\bibitem{2010:Belesi:PhysRevB}
	\bibinfo{author}{Belesi, M.} \emph{et~al.}
	\newblock \bibinfo{title}{{Ferrimagnetism of the magnetoelectric compound
			Cu$_{2}$OSeO$_{3}$ probed by $^{77}$Se NMR}}.
	\newblock \emph{\bibinfo{journal}{Phys. Rev. B}} \textbf{\bibinfo{volume}{82}},
	\bibinfo{pages}{094422} (\bibinfo{year}{2010}).
	\newblock \urlprefix\url{http://link.aps.org/doi/10.1103/PhysRevB.82.094422}.
	
	\bibitem{Janson2014}
	\bibinfo{author}{Janson, O.} \emph{et~al.}
	\newblock \bibinfo{title}{{The quantum nature of skyrmions and half-skyrmions
			in Cu$_{2}$OSeO$_{3}$}}.
	\newblock \emph{\bibinfo{journal}{Nat. Commun.}} \textbf{\bibinfo{volume}{5}},
	\bibinfo{pages}{5376} (\bibinfo{year}{2014}).
	\newblock \urlprefix\url{http://dx.doi.org/10.1038/ncomms6376}.
	
	\bibitem{2012:Adams:PhysRevLett}
	\bibinfo{author}{Adams, T.} \emph{et~al.}
	\newblock \bibinfo{title}{{Long-Wavelength Helimagnetic Order and Skyrmion
			Lattice Phase in Cu$_{2}$OSeO$_{3}$}}.
	\newblock \emph{\bibinfo{journal}{Phys. Rev. Lett.}}
	\textbf{\bibinfo{volume}{108}}, \bibinfo{pages}{237204}
	(\bibinfo{year}{2012}).
	\newblock
	\urlprefix\url{http://link.aps.org/doi/10.1103/PhysRevLett.108.237204}.
	
	\bibitem{2012:Seki:PhysRevB}
	\bibinfo{author}{Seki, S.} \emph{et~al.}
	\newblock \bibinfo{title}{{Formation and rotation of skyrmion crystal in the
			chiral-lattice insulator Cu$_{2}$OSeO$_{3}$}}.
	\newblock \emph{\bibinfo{journal}{Phys. Rev. B}} \textbf{\bibinfo{volume}{85}},
	\bibinfo{pages}{220406 (R)} (\bibinfo{year}{2012}).
	\newblock \urlprefix\url{http://link.aps.org/doi/10.1103/PhysRevB.85.220406}.
	
	\bibitem{2012:Seki:PhysRevB2}
	\bibinfo{author}{Seki, S.}, \bibinfo{author}{Ishiwata, S.} \&
	\bibinfo{author}{Tokura, Y.}
	\newblock \bibinfo{title}{{Magnetoelectric nature of skyrmions in a chiral
			magnetic insulator Cu$_{2}$OSeO$_{3}$}}.
	\newblock \emph{\bibinfo{journal}{Phys. Rev. B}} \textbf{\bibinfo{volume}{86}},
	\bibinfo{pages}{060403} (\bibinfo{year}{2012}).
	\newblock \urlprefix\url{http://link.aps.org/doi/10.1103/PhysRevB.86.060403}.
	
	\bibitem{2012:White:JPhysCondensMatter}
	\bibinfo{author}{White, J.~S.} \emph{et~al.}
	\newblock \bibinfo{title}{{Electric field control of the skyrmion lattice in
			Cu$_{2}$OSeO$_{3}$}}.
	\newblock \emph{\bibinfo{journal}{J. Phys. Condens. Matter}}
	\textbf{\bibinfo{volume}{24}}, \bibinfo{pages}{432201}
	(\bibinfo{year}{2012}).
	\newblock \urlprefix\url{http://stacks.iop.org/0953-8984/24/i=43/a=432201}.
	
	\bibitem{2013:Okamura:NatCommun}
	\bibinfo{author}{Okamura, Y.} \emph{et~al.}
	\newblock \bibinfo{title}{{Microwave magnetoelectric effect via skyrmion
			resonance modes in a helimagnetic multiferroic}}.
	\newblock \emph{\bibinfo{journal}{Nat. Commun.}} \textbf{\bibinfo{volume}{4}},
	\bibinfo{pages}{2391} (\bibinfo{year}{2013}).
	\newblock
	\urlprefix\url{http://www.nature.com/ncomms/2013/130830/ncomms3391/full/ncomms3391.html}.
	
	\bibitem{WhitePRL2014}
	\bibinfo{author}{White, J.~S.} \emph{et~al.}
	\newblock \bibinfo{title}{{Electric-Field-Induced Skyrmion Distortion and Giant
			Lattice Rotation in the Magnetoelectric Insulator
			${\mathrm{Cu}}_{2}{\mathrm{OSeO}}_{3}$}}.
	\newblock \emph{\bibinfo{journal}{Phys. Rev. Lett.}}
	\textbf{\bibinfo{volume}{113}}, \bibinfo{pages}{107203}
	(\bibinfo{year}{2014}).
	\newblock
	\urlprefix\url{http://link.aps.org/doi/10.1103/PhysRevLett.113.107203}.
	
	\bibitem{2015::Okamura}
	\bibinfo{author}{Okamura, Y.} \emph{et~al.}
	\newblock \bibinfo{title}{{Microwave Magnetochiral Dichroism in the
			Chiral-Lattice Magnet Cu$_{2}$OSeO$_{3}$}}.
	\newblock \emph{\bibinfo{journal}{Phys. Rev. Lett.}}
	\textbf{\bibinfo{volume}{114}}, \bibinfo{pages}{197202}
	(\bibinfo{year}{2015}).
	\newblock
	\urlprefix\url{http://link.aps.org/doi/10.1103/PhysRevLett.114.197202}.
	
	\bibitem{Mochizuki2015PRL}
	\bibinfo{author}{Mochizuki, M.}
	\newblock \bibinfo{title}{{Microwave Magnetochiral Effect in
			${\mathrm{Cu}}_{2}{\mathrm{OSeO}}_{3}$}}.
	\newblock \emph{\bibinfo{journal}{Phys. Rev. Lett.}}
	\textbf{\bibinfo{volume}{114}}, \bibinfo{pages}{197203}
	(\bibinfo{year}{2015}).
	\newblock
	\urlprefix\url{http://link.aps.org/doi/10.1103/PhysRevLett.114.197203}.
	
	\bibitem{2014::Yu::SciRep}
	\bibinfo{author}{Yu, H.} \emph{et~al.}
	\newblock \bibinfo{title}{{Magnetic thin-film insulator with ultra-low spin wave damping for coherent nanomagnonics}}.
	\newblock \emph{\bibinfo{journal}{Sci. Rep.}}
	\textbf{\bibinfo{volume}{4}}, \bibinfo{pages}{6848}
	(\bibinfo{year}{2014}).
	\newblock
	\urlprefix\url{http://dx.doi.org/10.1038/srep06848}.
	
	\bibitem{2013:AllivyKelly:APL}
	\bibinfo{author}{d'Allivy Kelly, O.} \emph{et~al.}
	\newblock \bibinfo{title}{{Inverse spin Hall effect in nanometer-thick yttrium iron garnet/Pt system}}.
	\newblock \emph{\bibinfo{journal}{Appl. Phys. Lett.}}
	\textbf{\bibinfo{volume}{103}}, \bibinfo{pages}{082408}
	(\bibinfo{year}{2013}).
	\newblock
	\urlprefix\url{http://dx.doi.org/10.1063/1.4819157}.
	
	\bibitem{Kalarickal::JAPL::2006}
	\bibinfo{author}{Kalarickal, Sangita S.} \emph{et~al.}
	\newblock \bibinfo{title}{{Ferromagnetic resonance linewidth in metallic thin films: Comparison of measurement methods}}.
	\newblock \emph{\bibinfo{journal}{J. Appl. Phys.}}
	\textbf{\bibinfo{volume}{99}}, \bibinfo{pages}{093909}
	(\bibinfo{year}{2006}).
	\newblock
	\urlprefix\url{http://dx.doi.org/10.1063/1.2197087}.	

	\bibitem{Giesen2005}
	\bibinfo{author}{Giesen, Fabian}
	\newblock \bibinfo{title}{{Magnetization Dynamics of Nanostructured Ferromagnetic Rings and	Rectangular Elements}}.
	\newblock \emph{\bibinfo{journal}{PhD thesis, University of Hamburg}}
	(\bibinfo{year}{2005}).
	
	
	
\end{thebibliography}

\end{document}